# Nanomaterials : a review of the definitions, applications, health effects. How to implement secure development

E. Gaffet

Nanomaterials Research Group – UMR CNRS 5060

Site de Sévenans (UTBM) – F90010 Belfort Cedex - France

Eric.Gaffet@utbm.fr

**Résumé :**

*Les nanomatériaux représentent un domaine de recherche actif mais aussi un secteur économique en pleine expansion en vue de nombreuses applications. Par exemple la production française pour les matériaux les plus courants (comme la silice, le dioxyde de titane, le noir de carbone) se chiffre en centaines de milliers de tonnes. Comme c'est le cas pour toute innovation, il convient de s'interroger sur les risques et, si nécessaire, de fixer des règles pour protéger la santé du consommateur et celle du travailleur. On discute en particulier les difficultés pour définir ces matériaux, de l'état des connaissances en matière de toxicité humaine ou environnementale et des prescriptions des agences en matière de sécurité.*

**Abstract:**

*Nanomaterials are an active area of research but also an economic sector in full expansion which addresses many applications domains. For instance, french production for the most common nanomaterials (such as silica, titanium dioxide, carbon black) is in the hundreds of thousands of tons. As for any innovation, one must consider the risks and, if necessary, establish rules to protect consumer health and that of the worker. This paper addresses in particular difficulties in defining these materials, the state of knowledge on human or environmental toxicity and requirements and agencies in charge of safety.*

## I. Introduction

Nanomaterials and nano-manufactured goods represent areas of scientific research and industrial applications in full expansion. They are already an industrial and economic reality. As for other industrial sectors, the introduction of these new products should be considered in terms of potential effects of toxicity and ecotoxicity, so to measure and to control as well the societal consequences that environmental and health.

## 2 Definition of the perimeter of nanomaterials

To define the concept of nanoparticles and nanomaterials and/or, it took until the pre - normative terminology such as defined by the ISO[1] published at the end of September 2008 under the number ISO / TS 27687[2] ("Nanotechnology - terminology and definitions for the "nano-objects", i.e. Nanoparticle, nanofiber and nanoplate"), namely :

(i) nano-definition considered the range of dimension between 1 and 100 nm.

(ii) nano - objects are materials with one, two or three external in the nano-domain dimensions. Among these nano-objects, nanoparticles include their three dimensions in the field of nano, nano - plates have one nano-sized dimension, nanofibers are nano - objects with two nano-sized dimensions, the third dimension being significantly longer.

The 2009's and 2010's can be considered as very important concerning the definition of the field of nanomaterials and/or nanoparticles. They are indeed reflected by various national bodies.

Thus, "Health Canada" (Ministry of Health in Canada) introduced a draft definition of the nanomaterials[3]. "Any manufactured product, material, substance, device, system or structure should be regarded as nanomaterials."
i) If it is at the nanoscale, or within it, in at least one dimension space,
ii) If it is smaller or larger than the nanoscale in all dimensions and displays a nano-scale phenomenon or more.
Health Canada States that the "nano-scale" means 1 to 100 nanometers, i.e. promoting ISO definition. However, two specific concepts are added. First of all, that of nano-scale phenomena which are "*attributable to its size and* distinguishable from the chemical or physical properties of individual atoms, individual molecules and bulk material".

On the other hand, "Health Canada" distinguished manufactured goods stating "the term "manufactured includes those technical processes and controls of matter and processes at the nanoscale".

Similarly, the NICNAS (The National Industrial Chemicals Notifications and Assessment Scheme is the Australian Government regulatory scheme for industrial chemical) defines the scope nano for entry into force on 1 January 2011[4] "... *industrial materials intentionally produced, manufactured, modified to have unique properties or a specific nano-scale structure.*", i.e. between 1 and 100 nanometers, and that is either a nano object (i.e. confined to this scale in one, two or three dimensions) or nanostructured (i.e. with an internal structure or surface at this scale). Once again, the 1-100 nm scale advocated by ISO is used. The NICNAS has the same concerns as Health-Canada. First it distinguishes materials intentionally produced, manufactured or modified from those produced accidentally. Second, it puts forward the ability of these objects to exhibit unique properties not found in the same non-nanostructured material.

---

[1] ISO : International Standard Organisation
[2] http://www.iso.org/iso/catalogue_detail?csnumber=44278
[3] *"Interim Policy Statement on Health Canada's Working Definition for Nanomaterials"*- http://www.hc-sc.gc.ca/sr-sr/consult/_2010/nanomater/draft-ebauche-eng.php
[4] *"Adjustments to NICNAS new chemicals processes for industrial nanomaterials"*
http://www.nicnas.gov.au/Publications/Chemical_Gazette/pdf/2010oct_whole.pdf#page=14

In addition, the NICNAS takes into account two extensions of the nano field: On the one hand aggregates or agglomerated materials are considered as nanostructured. On the other hand all materials of which the size of components distribution contains at least 10% of nano-scale particles are considered as nanomaterials.

In October 2010, the European commission launches a consultation[5] on the definition of Nanomaterials and, on this occasion, proposes a recommendation[6]:"*Member States, the European agencies and industry are invited to use this definition of nanomaterial when they adopt and implement of regulations or programs relating to the products of nanotechnologies*".

"*Nanomaterial: means a material that meets at least one of the following criteria:*
- *consists of particles, with one or more external dimensions in the size range 1 nm - 100 nm for more than 1 % of their number size distribution;*
- *has internal or surface structures in one or more dimensions in the size range 1 nm- 100 nm;*
- *has a specific surface area by volume greater than 60 $m^2/cm^3$, excluding materials consisting of particles with a size lower than 1 nm*".

The reading of these proposals for definitions - some even to the status of "draft" or simple recommendation - translates very clearly the taken into account by the authorities in charge of the framework for the development of these industrial sectors, of the specificities of nanomaterials, namely: the variation of physico-chemical properties, the importance of specific surface area, and the integration of the concept of size distribution of nanoparticles with consideration of a minimum content from which a product / component or a substance should be considered as falling within the field "nano". It should be noted the difficulty, for definitions and/or regulations, to define precise bounds in terms of dimensions for the field "nano". Thus, the dimensional limits 1 nm and 100 nm can be strict and exclusive or indicative.

In addition to these attempts of definition, some almost entered the common language. By extension (and without doubt, abuse of language with respect to this definition), it is very often customary to extend the notion of nanoparticles to the whole of the nano - objects, while disregarding the number of dimensions of the object below 100 nm. Moreover, in this review, nanomaterials are to be regarded as a more or less compact and dense assembly of nano - objects (aggregates and / or agglomerates).

---

[5] http://ec.europa.eu/environment/consultations/nanomaterials.htm
[6] http://ec.europa.eu/environment/consultations/pdf/recommendation_nano.pdf

## 3 State of industrial production and fields of applications

### 3.1. State of the production

The extremely rapid development of nanomaterials and manufactured nanoparticles involves potential exposure to manufactured nanomaterials of a more and more significant population, in particular industrial operators (near 3270 staffs in France for the production) or research staff laboratories (approximately 7000 staffs in France).

With regard to French industrial production of nanomaterials, the conclusion of a study[7] published in July 2008 by the French Agency for sanitary security of the environment and the workplace (AFSSET, which in the meantime became the ANSES i.e. the national agency for sanitary security), relying on a previous study[8] of the National Institute for research and safety (INRS) indicates the following annual productions in France:

- About 485.000 tons of silica by approximately 1300 operators[9],
- 469.000 tons of alumina by nearly 1,000 operators,
- 300,000 tons of calcium carbonate,
- 240.000 tons of black carbon by 280 operators,
- 250,000 tons form sub - micronic titanium dioxide and 10,000 tons as nanoparticles involving 270 operators,
- The annual French production capacity of carbon nanotubes is 10 tons involving 10 operators.

These figures for French production can be compared to the amount of these nanoparticles in the natural state. According to an article by Buzea and al.[10] anthropogenic aerosols (i.e. due to human activity) represent only 10% of the total of these nanoparticles in the atmosphere. The annual source, of naturally occurring nanoparticles worldwide is estimated to about 16.8 million tons for mineral aerosol due mainly to wind erosion. Note that 1% of mineral aerosol is produced in volcanic eruptions. To these mineral aerosols, one should add 3.6 million tons of marine salts and 1.8 million tons of nanoparticles from biomass.

---

[7] *"Nanomatériaux et Sécurité au Travail"*, published in July 2008
http://www.afsset.fr/upload/bibliotheque/258113599692706655310496991596/afsset-nanomateriaux-2-avis-rapport-annexes-vdef.pdf
[8] *"Production et utilisation industrielle des particules nanostructurées"*
HST ND 2277 - 209 – 07, paru en 2007 - Bertrand Honnert, Raymond Vincent
http://www.inrs.fr/inrs-pub/inrs01.nsf/IntranetObject-accesParReference/ND%202277/$File/ND2277.pdf
[9] Operators means the people how are potentially exposed at workplace
[10] *"Nanomaterials and nanoparticles: Sources and toxicity"*
C. Buzea, I.. I. Pacheco Blandino, K. Robbie - Biointerphases, vol. 2, issue 4 (2007) pages MR17 - MR172

The annual tonnage of sulphate of natural and anthropogenic origins is 3.3 millions tons. Industrial soot represents nearly 1.4 million tons. Nanoparticles of hydrocarbons (non-methane) are issued at a rate of 1.3 million tons per year. The annual tonnage of nitrate of natural and anthropogenic origins is 0.6 million ton. The tonnage of nanoparticles resulting of the natural process of degradation of biological organisms is also estimated to 0.6 million ton. Finally, these tonnages must be added to nearly 20,000 tons of cosmic dust

The importance of these non-anthropogenic emissions of nanoparticles should be noted because they come to constitute natural background noise, which we need to extract it to get the content of manufactured nanoparticles.

### 3.2. Scope of Applications

The applications of nanomaterials are multiple, as developed following according to the report entitled "Prospective study on nanomaterials", carried out on behalf of the Directorate General of industry, technologies of information and the Department of Finance (DIGITIP) positions by French development & Council (2004)[11].

---

[11] *"Etude prospective sur les Nanomatériaux"*
http://www.industrie.gouv.fr/enjeux/pdf/synthesenanomateriaux.pdf

| Nanomaterials | Application fields |
|---|---|
| **Nanoceramics** | Structural composite materials - components anti - UV - polishing substrates (wafers) in microelectronics Chemical - photocatalytical applications |
| **Nanometallics** | Antimicrobial and/or sector of catalysis - conductive layers of screens, sensors or energetic materials. |
| **Nanoporous** | Aerogels for thermal insulation in the areas of electronics, optics and catalysis - bio - medical field for tracing or even implants type applications. |
| **Nanotubes** | Electrical conductive nanocomposites - structural materials - single-walled nanotubes for applications in the field of electronics, screens |
| **Massive Nanomaterials** | Hard coatings - structural components for the aerospace industry, automotive, pipes for oil and gas, sport or even anticorrosion sector industries. |
| **Dendrimers** | Medical field (administration of drugs, rapid detection) - domain cosmetic. |
| **Quantum Dots** | Optoelectronics (screens) - photovoltaic cells - inks and paints for applications of type marking anti - counterfeiting |
| **Fullerenes** | Sport (nanocomposites) and cosmetics sectors |
| **Nanowires** | Applications in the conductive layers of screens or even solar cells and electronic devices |

**Table 1:** Fields of applications by type of nanomaterials

In the food sector, the National Agency for Sanitary Security - France (ANSES – Agence Nationale de Sécurité Sanitaire) has included in a report published in 2010[12], a case study of the use of silica nano as anti-caking for powdered sugar. A recent study by the National Institute for public health and the environment, the equivalent of the ANSES to the Netherlands allows to grasp concretely the magnitude of the very broad use of nanoparticles in silica as additives in non-negligible percentages, for example for type anti-caking applications, anti - agglomerating or even as a viscosity modifier: sauce for lasagna, instant noodles, various seasoning for meat and burrito, pancake, pepper, coffee cream, roasted vegetables, lasagna sauce mix[13].

---

[12] *"Les Nanomatériaux : Évaluation des risques liés aux nanomatériaux pour la population générale et dans l'environnement"* - Saisine Afsset n° 2008/005"
F. Nesslany, M. Boize, J.-Y. Bottero, D. Chevalier, E. Gaffet, O. Le Bihan, C. Mouneyrac, M. Riediker, F. Tardif - Rapport Agence Française de Sécurité Sanitaire, de l'Environnement et du Travail, Mars 2010.
http://lesrapports.ladocumentationfrancaise.fr/BRP/104000168/0000.pdf
[13] *"Presence and risks of nanosilica in food products"*
S. Dekkers, P. Krystek, R. J.B. Peters, D. P.K. Lankveld, B.G.H. Bokkers, P.H. Van Hoeven-Arentzen, H. Bouwmbester, A.G. Oomen - Nanotoxicology J 2010; Early Online J 1-13

## 4 State of toxicity and EcoToxicité knowledge

The State of knowledge on the effects of micro/nano particles atmospheric pollution makes fear effects on health of man - produced[14,15] nanoparticles. Nevertheless, very little reliable data - in the sense of reproduced by different research teams - are currently available in this area. However the published studies which report interactions of nanoparticles with the living at the cellular level, encourage caution[16].

In the field of toxicology and/or eco - toxicology research, the effects of particles are deemed be correlated to the mass of product to which animals or humans are exposed: the greater the absorbed mass, the greater is the effect.

In the case of nanoparticles, it has been clearly demonstrated[17,18,19] that the mass of the product is not the most relevant parameter to correlate with measured effects, thus upsetting the traditional interpretation of the measures of toxicity. It is indeed found, that equal mass, nanoparticles are more toxic that the products of the same chemical composition presented in fragments of larger size.

Although several studies have found a correlation between the specific surface area and the toxicity, there seems to be a consensus in the scientific community to agree that several factors contribute to the toxicity of these new generation products and that it is currently impossible, from our fragmentary knowledge, to balance their respective weight or accurately predict the toxicity of a new nanoparticle. Published studies link observed effects which are due to different parameters : specific surface area, number of particles[20], size and size distribution, concentration, surface state[21] (contamination, electric charge), degree of particle agglomeration and site of deposition in the lungs[22], the surface electrical charge, shape, porosity, crystalline structure, the potential of electrostatic attraction, method for particle synthesis, hydrophilic/hydrophobic character and post-synthesis modifications (the surface coating to prevent agglomeration). The presence of contaminants such as metals can also promote the formation of free radicals and inflammation. Similarly the particle chemical composition and leaching in the body of components present on its surface, the colloidal properties and surface of nanoparticles, compartmentalization in the respiratory tract and the bio - persistence are factors rendering more complex the understanding of their toxicity. In addition, the slow dissolution of some of nanoparticles or some of their components in the body may become a major element determining their toxicity. These factors will therefore influence their functional, toxicological and environmental impact.

---

[14] *"Characteristics of a Residential and Working Community With Diverse Exposure to World Trade Center Dust, Gas, and Fumes"* J. Reibman, M. Liu, Q. Cheng, MS, S. Liautaud, L. Rogers, S. Lau, K. I. Berger, R. M. Goldring, M.I Marmor, M. E. Fernandez-Beros, E. S. Tonorezos, C. E. Caplan-Shaw, J. Gonzalez, J. Filner, D. Walter,, K. Kyng, W. N. Rom - J Occup Environ Med. 2009 May; 51(5): 534– 541.
[15] *"A review of commuter exposure to ultrafine particles and its health effects"*
L. D. Knibbs , T. Cole-Hunter L. Morawska - Atmospheric Environment 45 (2011) 2611e2622
[16] *"Les Nanoparticules. Un enjeu majeur pour la santé au travail ?"*
sous la direction de Benoit Hervé – Bazin (INRS), EDP Sciences 2007
[17] *"Airborne nanostructured particles and occupational health"*
A. D. Maynard, E. D. Kuempel - Journal of Nanoparticle Research (2005) 7: 587–614
[18] *"Airborne nanostructured particles and occupational health"*
G. Oberdörster, E. Oberdörster, J. Oberdörster - Environ Health Perspect. 2005 Jul;113(7):823-39
[19] *"Concepts of nanoparticle dose metric and response metric"*
G. Oberdörster, E. Oberdörster, J. Oberdörster- Environ Health Perspect. 2007 June; 115(6): A290
[20] *"In search of the most relevant parameter for quantifying lung inflammatory response to nanoparticle exposure: particle number, surface area, or what?"*
K. Wittmaack.- Environ Health Perspect. 2007 Feb;115(2):187-94
[21] *"Skin penetration and kinetics of pristine fullerenes (C60) topically exposed in industrial organic solvents"*
Xin R. Xia, Nancy A. Monteiro-Riviere, Jim E. Riviere - Toxicology and Applied Pharmacology 242 (2010) 29–37
[22] *"Quantitative biokinetic analysis of radioactively labelled, inhaled Titanium dioxide Nanoparticles in a rat model"*
W. G. Kreyling A. Wenk, M. Semmler-Behnke - http://www.umweltdaten.de/publikationen/fpdf-l/4022.pdf

The toxicological effects of nanomaterials began to be investigated in recent years on animal models and cell culture. There are almost no studies on toxicological effects of nanomaterials for humans. Due to the small number of studies, the advent of recent nanomaterials and their variability, interpretation of experimental studies on the effects of nanomaterials requires great caution.

Among various mechanisms envisaged to understand specific toxicity of nanoparticles, a recent publication describes mainly two major types of mechanisms[23] as being at the origin of the toxicological effects of nanomaterials:

(i) oxidative stress induction: metal nanomaterials can produce significant quantities of reactive forms of oxygen (still called free radicals) on their surface or induce their excessive production by cells. These molecules are of a high biological reactivity, and can damage the walls of the cells, which can induce reactions of inflammation, and fibrosis. They can also adversely affect the DNA of cells and favor cancerous process.

(ii) surface adsorption of biologically active molecules:, because of their specific surface properties, nanomaterials may adsorb biologically active molecules, as factors required for cell growth, thus inducing cell suffering and sometimes death, be nanoparticles internalized in cells or not[24]. These effects have been well documented for the lung case[25]. Moreover, according to studies with animal models, exposure to some particular carbon nanotubes can induce damage similar to those induced by asbestos[26][27].

---

[23] *"Nanoparticules : une prévention est-elle possible ? - Nanoparticles : Is a prevention possible?"*
J. Boczkowski, S. Lanone - Revue française d'allergologie Volume 50, numéro 3 pages 214-216 (avril 2010)
[24] *"Engineered nanoparticles Rev Health Environmental Safety"*
Aitken R, Aschberger K, Baun A, Christensen F, Fernandes TF, Hansen S, Hartmann N, Hutchison G, Johnston H, Micheletti C- Enrhes (2010)
[25] *"Mechanisms of pulmonary toxicity and medical applications of carbon nanotubes : two faces of janus ?"*
Shvedova AA, Kisin ER, Porter D, Schulte P, Kagan VE, Fadeel B, Castranova V. Pharmacol Ther 2009; 121 : 192-204.]
[26] *"Induction of mesothelioma in p53+/- mouse by intraperitoneal application of multi-wall carbon nanotube"*
Takagi A, Hirose A, Nishimura T, Fukumori N, Ogata A, Ohashi N, et al.. J Toxicol Sci 2008; 33: 105-16,
[27] *"Carbon nanotubes introduced into the abdominal cavity of mice show asbestoslike pathogenicity in a pilot study"*

For animals, as a consequence of the two mechanisms described above by S Lanone and J. Boczkowski, several effects have already been demonstrated including toxic effects for various organs (heart, lungs, kidneys, reproducing organs...), as well as genotoxcity[28,29] (DNA damage) and cytotoxicity (cell toxicity)[30]. Some particles, for example, cause granulomas, fibrosis and tumor reactions at chest level. Thus a substance reputed to be little harmful, titanium dioxide, exhibits significant pulmonary toxicity when it is nano-sized.

Overall, nanoparticle-specific toxicological data remain limited. Beyond very partial data, quantitative assessment of risk is difficult for most substances, in particular because of short periods of exposure, of various compositions of tested nanoparticles (diameter, length and agglomeration). in addition the route of exposure used for toxicity assessments is often not representative of occupational exposure. Additional studies (absorption, bio - persistence, carcinogenicity, translocation to other tissues or organs, accumulation, etc.) are therefore needed to dispose of all of the information required for the quantitative assessment of the risk associated with exposure by inhalation, percutaneous exposure among workers. Available data clearly indicate that certain insoluble nanoparticles can cross different biological protection barriers, to disseminate inside the body and accumulate in some organs and cells. Toxic effects have already been documented for lungs, heart, reproductive system, kidney, skin and at cellular level. Significant accumulations were reported in lungs, brain, liver, spleen, and bone. Among biological barriers which nanoparticles cross easily, the placental barrier[31,32] should be noted

In the light of this brief summary, embodying the elements of a synthesis of the work of literature synthesis published by the Institut de Recherche Robert-Sauvé in health and safety at workplace in Quebec (IRSST) in April 2008[33] or those from a australian study[34], major trends are emerging and reveal many toxic effects related to certain nanoparticles.

---

C. A. Poland, R. Duffin, I. Kinloch, A. Maynard, W. A. H. Wallace, A. Seaton, V. Stones, S. Brown, W. Macnee, K. Donaldson, Nature Nanotechnology, Published online: 20 May 2008; doi:10.1038/nnano.2008.111]

[28] *"Titanium Dioxide Nanoparticles Induce DNA Damage and Genetic Instability In vivo in Mice"*
B. Trouiller, R. Reliene1, A. Westbrook, P. Solaimani1, R. H. Schiestl - Cancer

[29] *"Nanoparticles can cause DNA damage across a cellular barrier"*
G. Bhabra, A. Sood, B. Fisher, L. Cartwright, M. Saunders, W. H. Evans, A. Surprenant, G. Lopez- Castejon, S. Mann, S. A. Davis, L. A. Hails, E. Ingham, P. Verkade, J. Lane, K. Heesom, R. Newson, C. P. Case
Nature Nanotechnology 4, 876 - 883 (2009)

[30] *"SiO$_2$ nanoparticles induce cytotoxicity and protein expression alteration in HaCaT cells"*
X. Yang, J. Liu, H. He, L. Zhou, C. Gong, X. Wang, L. Yang, J. Yuan, H. Huang, L. He, B. Zhang -
Particle and Fibre Toxicology, 2009, 7: 1 (42pp)

[31] *"Transfer of Quantum Dots from Pregnant Mice to Pups Across the Placental Barrier"*
M. Chu, Q. Wu , H. Yang, R. Yuan, S. Hou, Y. Yang, Y. Zou, S. Xu, K. Xu, A. Ji, L. Sheng - Small – 2010 Volume 6 Issue 5, Pages 670 - 678

[32] *"Effects of prenatal exposure to surface-coated nanosized titanium dioxide (UV-Titan). A study in mice"*
K. S Hougaard, P. Jackson, K. A Jensen; J. J Sloth, K. Loschner , E. H Larsen, R. K Birkedal, A. Vibenhol , AM Z Boisen, H. Wallin, U. Vogel - Particle and Fibre Toxicology 2010, 7:16 doi:10.1186/1743-8977-7-16

[33] *"Les effets sur la santé reliés aux nanoparticules, seconde édition"*

It is clear that each synthesized product can have its own toxicity. This implies then studies "on a case by case" basis, and especially very great caution in extrapolating, or even generalizing the results of tests on one or other of various nanoparticles to all for example the same chemical composition or same dimension nanoparticles.

On the basis of the work published to date, it seems indeed that any modification of the process used to produce a given nanoparticle or to modify its surface could have an impact on its toxicity. Here, it should be noted that changes of surface properties for a given particle type will depend on the life cycle of the material. Consider the example of titanium dioxide. It is exposed to different environments (hence to different contaminant) depending on its implementation in a solar cream, embedded in a varnish or introduced into concrete.

This analysis of the IRSST is also supported by a journal article by Hansen et al.[35] on the basis of the analyses of 428 published studies about 965 nanoparticle toxic effects, it appears that on these 428 studies, 120 indicate a specific toxicity in mammals and 270 a cytotoxicity in "in vitro" experiments. The journal article highlights also the variability of nanomaterials to be considered and little information actually available (and relevant) on the nature of nanoparticles studied.

This observation of the importance of the variability of physic-chemical parameters on the results of the tests to study the toxicity and ecotoxicity of these nanoparticles led the Working Party on Manufactured NanoMaterials (WPMN) work carried out at the level of the OECD to clarify the critical parameters to set. Physico - chemical parameters and toxicity and ecotoxicity tests to perform to assess the dangerousness of nanoparticles are listed below.

---

Études et recherches / Rapport R-558, Montréal, IRSST, paru en Avril 2008. 1) Claude Ostiguy, Brigitte Soucy, Gilles Lapointe, Catherine Woods et Luc Ménard http://www.irsst.qc.ca/files/documents/PubIRSST/R-558.pdf *"Les nanoparticules : connaissances actuelles sur les risques et les mesures de prévention en santé et en sécurité du travail"* and 2) Claude Ostiguy, Gilles Lapointe, Luc Ménard, Yves Cloutier, Mylène Trottier, Michel Boutin, Monty Antoun, Christian Normand Études et recherches / Rapport R-455, Montréal, IRSST, paru en 2006 http://www.irsst.qc.ca/files/documents/PubIRSST/R-455.pd
[34] *"Engineered Nanomaterials: A review of the toxicology and health hazards"*
Safe Work Australia (2009) –
http://www.safeworkaustralia.gov.au/AboutSafeWorkAustralia/WhatWeDo/Publications/Documents/313/EngineeredNanomaterials_%20Review_ToxicologyHealthHazards_2009_PDF.pdf
[35] *"Categorization framework to aid hazard identification of nanomaterials"*
S. F. Hansen, B. H. Larsen, S. I. Olsen, A. Baun - Nanotoxicology, 1 - 8 (2007)
http://dx.doi.org/10.1080/17435390701727509

| Physical-Chemical Properties and Material Characterization | Environmental Fate |
|---|---|
| • Agglomeration/aggregation<br>• Water solubility<br>• Crystalline phase<br>• Dustiness<br>• Crystallite size<br>• Representative TEM picture(s)<br>• Particle size distribution<br>• Specific surface area<br>• Zeta potential (surface charge)<br>• Surface chemistry (where appropriate)<br>• Photocatalytic activity<br>• Pour density<br>• Porosity<br>• Octanol-water partition coefficient,<br>• Redox potential<br>• Radical formation potential<br>• Other relevant information | • Dispersion stability in water<br>• Biotic degradability<br>• Ready biodegradability<br>• Simulation testing on ultimate degradation in surface water<br>• Soil simulation testing<br>• Sediment simulation testing<br>• Sewage treatment simulation testing<br>• Identification of degradation product(s)<br>• Further testing of degradation product(s) as required<br>• Abiotic degradability and fate<br>• Hydrolysis, for surface modified nanomaterials<br>• Adsorption- desorption<br>• Adsorption to soil or sediment<br>• Bioaccumulation potential<br>• Other relevant |
| **Environmental Toxicology**<br>• Effects on pelagic species<br>• Effects on sediment species<br>• Effects on soil<br>• Effects on terrestrial species<br>• Effects on microorganisms<br>• Other relevant information<br>**Mammalian Toxicology**<br>• Pharmacokinetics (ADME)<br>• Acute toxicity<br>• Repeated dose toxicity<br>If available:<br>• Chronic toxicity<br>• Reproductive toxicity<br>• Developmental toxicity<br>• Genetic toxicity<br>• Experience with human exposure<br>• Other relevant test data | **Material Safety**<br>Where available:<br>• Flammability<br>• Explosivity<br>• Incompatibility |

**Table 2:** List of critical physico - chemical parameters and nature of tests to determine the dangerousness of nanoparticles of interest (OECD / WPMN)[36]

As indicated in the IRSST 2008 report cited above, in the case of soluble nanoparticles, toxicity is related to their composition, regardless of their original size. After dissolution of the nanoparticles, toxicity is purely chemical origin. According to the same report, the situation is quite different for insoluble or very little soluble nanoparticles; size plays a role in this case.

Data currently available on toxicity of insoluble nanoparticles are fragmentary and do not always allow a quantitative risk assessment (or extrapolation to humans for nanoparticles synthesis),

---

[36] *"Series on the safety of manufactured nanomaterials number 6 list of manufactured nanomaterials and list of endpoints for phase one of the oecd testing programme"* ( ENV/JM/MONO(2008)13/REV (July 2008) - http://www.olis.oecd.org/olis/2008doc.nsf/LinkTo/NT000034C6/$FILE/JT03248749.PDF

except perhaps for TiO[37]. Elements of knowledge on their impact on the environment are even more limited. These elements on the concept of bio - persistence of insoluble nanoparticles and their health impact remain to this day relatively little explored both on animal models and on the environment in the broad sense

Given the importance and the activity in this area, various internet sites now allow the following of publications and other scientific work. This is the case for "The Virtual Journal on Nanotechnology, Environment, Health and Safety"[38] maintained by Rice University in the United States.

## 5 Ways to prevent

*Given these various findings and uncertainties about the health effects of nanoparticles, the AFSSET recommended in 2008 to declare nanoparticles as "level of unknown danger" and <u>manipulate them with the same caution as hazardous materials</u>*, i.e. to apply the safety procedures which are implemented to reduce the exposure to hazardous materials. Of course, as noted by the IRSST in conclusion of its report, the risk assessment should also take account of the route of exposure, its duration and the nanoparticles concentration and the susceptibility of the individual as well as the interaction of particles with biological components and biological fate.

The majority of the reports indicate that, given the lack of knowledge on the toxicity of nanoparticles, one is led to conclude that <u>the risk control is based on the control of exposure</u>. This point is developed in the next section which addresses the ways and methods of prevention i.e. "Nano - safety" as developed in the AFSSET [37] report.

At this point, two opinions of the High Council for Public Health in France (HCSP – Haut Conseil de la Santé Publique) should be stressed :
(i) The first one, which was published in January 2009[39], deals with carbon nanotubes. It builds on the report of the group in charge of following up research on the health impact of nanotechnology (created by the HCSP in 2008), on the Afsset's one (report of July 2008 quoted above), as well as on the advice[40] of the Committee of prevention and precaution published in May 2006. The High Council of public health says it considers the results of this expertise "*as constituting a major warning sign for the rapid implementation of measures to protect against exposures that might induce a serious health risk for producers and users of carbon nanotubes.*"

---

[37] *"NIOSH CURRENT INTELLIGENCE BULLETIN: Evaluation of Health Hazard and Recommendations for Occupational Exposure to Titanium Dioxide"*
National Institute for Occupational Safety and Health, published in December 2005
http://www.cdc.gov/niosh/review/public/Tio2/pdfs/TIO2Draft.pdf].
[38] http://www.icon.rice.edu/virtualjournal.cfm
[39] *"AVIS relatif à la sécurité des travailleurs lors de l'exposition aux nanotubes de carbone "*
HCSP - 7 janvier
2009 - http://www.hcsp.fr/docspdf/avisrapports/hcspa20090107_ExpNanoCarbone.pdf
[40] "*Nanotechnologies, nanoparticules. Quels dangers, quels risques ?"*
http://www.ladocumentationfrancaise.fr/rapports-publics/074000286/index.shtml

The High Council of Public Health (France) noted that nanoparticles should be subject to a procedure for registration and evaluation as of new chemical substances in Europe. He recommends to apply the precautionary principle, with respect to carbon nanotubes, i.e., to ensure that their production and their use (including for research activity) are under strict confinement conditions.

Moreover, the High Council of public health highlights the following actions it considers priority:

*1. Recommendations in terms of identification of situations of potential exposures and of nanosafety especially for carbon nanotubes:*

- *1.1 Identify and inventory worker populations likely to be exposed at each stage of the carbon nanotube lifecycle, in order to the implement their health surveillance. This provision will be facilitated by the duty of Declaration of use of nanoparticles that the HCSP wishes to be enacted, as expected, in the framework of the Grenelle act on environment, as well as by the duty of labeling products containing nanoparticles also provided for in this framework.*
- *1.2 Develop information of people affected by the risks and their prevention and associated training. These actions should particularly target the committees and the teams in charge of of occupational medicine. This means including the markup of the premises and / or the work stations affected by the use of carbon nanotubes (pictogram for nano-hazard)*
- *1.3 Include in the "safety" folder of any producer or user of carbon nanotubes a "nano-safety" component, which takes into account all of the life cycle of nanomaterials in the establishment, since their introduction in the workplace until their elimination. This provision should also apply to companies in charge of nanomaterial waste disposal.*
- *1.4 For substances containing carbon nanotubes, request that the safety data sheet explicitly mentions the presence of carbon nanotubes, as well as the need to implement precautionary practices.*
- *1.5 Organize traceability of exposure, through an individual exposure sheet as it is defined in the labor code, intended to be part of the medical record of each potentially exposed person.*

*2 Recommendations for research on metrology and risks*

*The High Council of public health recommends the development of a research priority in the following areas:*

- *2.1 Metrology of exposure to carbon nanotubes and other nanoparticles. As soon as possible it must be associated with the constitution a database about exposures.*
- *2.2 Evaluation and improvement of processes involving at different stages, carbon nanotubes and of collective and individual protection equipment.*
- *2.3 Constitution of a cohort of workers potentially exposed to nanoparticles, the priority being to this day to carry out a census of the people concerned and the compendium all relevant data to characterize their potential exposure.*

*2.4 Development of biological indicators of exposure and markers for early effects, in order to allow, where appropriate, the implementation of adapted medical follow-up.*

*2.5 Experimental toxicology for assessment of toxicity and carcinogenicity of nanotubes depending on the different routes of exposure. Currently, it is important to investigate the role of nanotube size, shape, rigidity and of the bio-persistence, including the translocation capability.*

*More generally, the High Council of the public health, aware that nanoparticles exhibit a wide range of physico-chemical characteristics and therefore, presumably, also of toxicological properties, recommends that are strengthened vigilance and research the possible health effects of different forms of nanoparticles.*

(ii) Tthe second notice of the HCSP published in March 2010 focused on silver nanoparticles[41]. The specialized Commission for risks - environment (CSRE) of the High Council of public health, noting wew silver nano structural forms and their new uses in varous goods and industrial applications, drew the attention of health authorities :

1. *On the need to implement a monitoring of the use of silver nano device in goods, especially those who come directly into contact with the consumer. Such a provision would ensure the traceability and information on the presence of nano silver, from production to the placing on the market, the disposal or recycling.*
   *This requires labelling of products and inventories (database) and must be accompanied by an assessment of the life cycle of the nano silver they contain.*
2. *On the need for research in particular on (i) the measurement of nano silver as well as its fate in food, water, air and (ii) a better understanding of its effects on man and the environment, including the consequences of the genotoxic and pro-apoptotic effects that it could lead to long term in connection with the different chemical structures of nano silver in consumer products.*
3. *On the importance of a toxicological and environmental assessment prior to introduction on the market for new products without waiting for implementation of EU regulation that French authorities are invited to promote, according to the founding principles of REACH and biocides regulations.*

The CSRE / HCSP invites health authorities to seize the competent health security agencies to ensure a scientific intelligence and an assessment of the risk. It declares itself ready, on this basis, to conduct an analysis on the options for risk management related to the use of nano silver in consumer products. More generally, it wishes risk/benefit assessments to be developed in the field of nanotechnologies applications leading to consumer's exposure.

---

[41] *"Recommandation de vigilance relative à la sécurité des nanoparticules d'Argent"*
Haut Conseil de la Santé Publique- 12 mars 2010
http://www.hcsp.fr/docspdf/avisrapports/hcspa20100312_nanoag.pdf

**5.1. Nano-safety (the so - called "STOP" principle)**

Developed and implemented means of prevention for nanoparticles follow the priority order classically used for chemical risk prevention. They rely on the STOP principle, as presented recently in the report of the AFSSET "Nanomaterials and occupational at worplace safety" published in July 2008.

The principle is to apply hierarchical control strategies. Protective measures should be defined and if necessary adapted to the results of the risk assessment. The strategy for monitoring protective measures must follow the "STOP" principle (Substitution, Technologies, Organization, personal Protection). This acronym indicates the priority order for these four axes of control. The first choice is the **S**ubstitution, followed by the **T**echnological measures and then of **O**rganizational measures. The need for individual **P**rotection should be avoided as much as possible. It applies only as an additional measure to reduce the risk, complementary to the other three strategies

● **Substitution**

Substition means:
- Replace toxic substances (basic particle material) by less toxic substances,
- Change the physical nature of the material,
- Change the type of application: this approach relates to the replacement of an application in powder or liquid spray (aerosol formation) by an application in liquid phase,
- Eliminate nanoparticles as soon as they are no longer necessary,
- Optimize equipments and processes,
- Favour forms which are non-dispersible in atmosphere, including suspensions in liquid or masters-mixtures.

● **Technologies**

This means implementing technical protection measures that are designed, as well as possible, to establish a barrier between workers and potentially dangerous substances or processes. Thus one effectively eliminates hazard exposure. The following approaches should be evaluated:
- Use closed systems,
- Use unbreakable containers or double containers for storage and transport,
- Manufacture and use the substance on a form which limit its dispersion,
- Capture pollutants at emitting source,
- Air Filtering prior to release,
- Separate work premises and adapt ventilation of the premises,
- Waste.

● **Organization**

Organizational measures were designed to reduce as much as possible interactions of staff with nanomaterials. These include different elements such as the limitation of the number of persons entering areas with probable exposures. Personnel only required for the current task and who have completed a safety training can enter. The staff control is ensured by a badge or a similar system. Moreover personnel working in high risk areas must be monitored (for example with device like a dead man switch). In addition, specific prevention and protection techniques should be considerd when one handles nanoparticles presenting a special risk of ignition or explosive (for example nano carbon or nano aluminum). This requires in particular the signaling of the danger of the area.

● **Individual Protection**

The so-called protective measures are the last line of defense against the occupational hazards. They include generally special clothing, protection against inhalation (mask), possibly overpressure of external air at the production site, skin protection, eye protection, protection against ingestion. One must take into account that these protections are often discomfortable or even constitute a significant physical load. They are sometimes difficult to apply or to wear properly. Individual protection measures will be implemented only if all other measures are not sufficient to achieve a level of acceptable risk. In this case, individual protective measures are a complement (not replacement) of the other measures.

### 5.2. Metrology - measurement and thresholds limits

In order to assess the operator residual exposure to nanomaterials, one must measure the atmosphere of the premises or the personnel exposure, according to assessed hazards, using available instrumentation. Measurements must distinguish manufactured nanomaterials from ambient background noise (natural particles and other sources). In cases where some characterizations are technically impossible, samples may be taken following determined protocols to be analyzed later when technical means will be available.

### 5.2.1. Means of action

Equipments that can be used to characterize the staff exposure are described in the annex to the AFSSET report of July 2008. They are ranked by increasing selectivity to particles with respect to background noise. We first find devices which measures only the particulate concentration, therefore, without any selectivity. Then we find equipments measuring both concentration and size (they discriminate particles of interest if these differ by size from the backgroud ones), and finally specific measuring methods (discrimination by nanoparticles composition or shape).

There is already a significant commercial offer to characterize the atmosphere of exposure to nanoparticles. These devices are based on principles of measurement sometimes very different:

electric charge, scattered light, elementary mass, etc. that lead to different measurement methodologies from those listed above - before.

In practice, however, note that it is not uncommon to obtain substantially different values of nanoparticles concentration for the same sample using two equipments yet based on the same principle. An important work of consolidation of the results obtained by the different methods of measurement and the development of simple calibration devices transportable is therefore necessary. This is the object of standardization work in the ISO context.

### 5.2.2 Recommendations for thresholds limits

In the particular case of carbon nanotubes, the National Institute for research and of security (INRS - France) stresses that the toxicological knowledge is still insufficient to establish values limits of occupational exposure to carbon nanotubes. He advocates that in the absence of limit value for exposure in French and European legislation, level of exposure as low as possible must be sought

In the case of $TiO_2$ nanoparticles, in the United States, the National Institute for safety to the work and health (NIOSH), expressed a specific risk for nanoparticles of $TiO_2$ proposing a threshold of 0.1 mg/m$^3$ for exposure value, i.e. a decrease by a factor of 15 with respect to the threshold value for micronic particles of the same compound. This threshold of 0.1 mg/m$^3$ corresponds to approximately $4.10^5$ particles of 50 nm per cm$^3$ of air or $5.10^4$ particles of 100 nm per cm$^3$. Thus, for a compound yet deemed to be weakly toxic in classical form, the proposed threshold corresponds to a number of nanoparticles that may be of the same order of magnitude as the background noise of natural particles. It is reasonable to think that even more severe thresholds will be awarded for more toxic nanoparticles than the $TiO_2$ ones.

The guide[42] of good practices of the British Institute of Standardization (BSI) proposes four types of nanoparticles and limit values for exposure based on those established limits currently for the larger particles. In some cases, exposure limit values are set at a lower level, since a material in nanoscale form may be more dangerous than a microscale one.

**i) Fibrous materials :** The BSI guide suggests as limit value exposure occupational fibers[43] the value of 0.01 fiber/ml, counts being carried out in scanning or transmission electron microscopy. This value is identical to the legal value after an asbestos site (counting in optical contrast of phase microcopy).

**ii) Nanomaterials based on carcinogenic, mutagenic or toxic substances for the reproduction :** Considering that, in nanoscale form, these materials may have a solubility and therefore an increased bioavailability, the guide suggests to adopt limit values for occupational exposure ten times lower than those in force now for these substances.

**iii) Insoluble nanomaterials :** The BSI guide endorsed the NIOSH work and proposals for recommendations (it is for the time being that of a working document) about titanium dioxide i.e. 1.5

---

[42] *"Nanotechnologies – Part 1: Good practice guide for specifying manufactured nanomaterials"*
http://www.bsi-global.com/upload/Standards%20&%20Publications/Nanotechnologies/PD6699-1.pdf
[43] Defined by a length greater than 5 micrometers and a length/width ratio > 3

mg/m$^3$ for fine particles but 0.1 mg/m$^3$ for ultrafine ones. Although it has been noted above that concentrations expressed in terms of mass are not necessarily representative of toxic effects, it is used still, until that other methods of measurement to be validated. In the absence of other studies, the BSI proposes that a similar approach applies to other insoluble nanomaterials, weighting existing limit values for occupational exposure by a factor of 0.066 (i.e. as for titanium dioxide). Another proposal for a limit value is based on the fact that urban pollution in the UK reaches commonly values lying between 20 000 and 50 000 nanoparticles per milliliter. The lower limit of 20 000 particles by milliliter (distinguishing ambient pollution particles) could be a value reference appropriate for insoluble nanoparticles.

**iv) Soluble nanoparticles:** for materials highly soluble regardless of their form, it is unlikely that nanoparticles present a superior bioavailability to those of larger particles. Similarly, it is unlikely they induce effects associated with insoluble nanoparticles. Therefore, it is proposed by the BSI to apply the values of occupational exposure limits a safety factor of 0.5.

**6 Conclusion**

All of the currently available studies, in vivo and in vitro, highlight the existence biological effects of nanomaterials in terms of inflammation, functional or structural, modulations at cellular level or for whole body. However, little data are currently available, and it seems urgent to deepen existing knowledge on the mechanisms involved in the dispersal of nanomaterials in the body. The risk assessment must not only take into account the potential effects of native nanomaterials, but also their behaviors and all of their life cycle (mode of manufacture, use, aging, bio-degradability...).

Moreover, it is essential to know levels and situations of exposure, and therefore the conditions of manufacture and composition of the products containing nanomaterials. In the absence of regulatory obligation, industrialists are very reluctant to communicate this information. In addition, epidemiological surveillance of workers exposed to nanomaterials should contribute decisively to the improvement of knowledge about their health effects possible medium and long-term.

Nanoscience (nanotechnologies and/or nanomaterials) are already an industrial and economic reality. Like other industrial sectors, should be questioned the introduction of these new products to measure and to control the consequences both societal that environmental and health. It's a crucial issue for the responsible and secure development of nanomaterials and nanotechnology.